\documentclass{article}

\usepackage{arxiv}

\usepackage[utf8]{inputenc} 
\usepackage[T1]{fontenc}    
\usepackage{url}            
\usepackage{booktabs}       
\usepackage{amsfonts}       
\usepackage{nicefrac}       
\usepackage{microtype}      
\usepackage{lipsum}
\usepackage{graphicx}
\usepackage{orcidlink}
\usepackage{comment}
\definecolor{VUB_blauw}{rgb}{0.1529, 0.2667, 0.5529}
\usepackage{hyperref}       
\hypersetup{
    colorlinks,%
    citecolor=VUB_blauw,%
    filecolor=VUB_blauw,%
    linkcolor=VUB_blauw,%
    urlcolor=VUB_blauw
}
\graphicspath{ {./images/} }

\usepackage[table,xcdraw]{xcolor}

\usepackage[most]{tcolorbox}
\usepackage{enumitem}
\usepackage[style=apa, backend=biber]{biblatex}
\AddToHook{shipout/background}{%
  \ifnum\value{page}=1 
    \pagenumbering{Roman} 
    \setcounter{page}{1} 
  \fi
  \ifnum\value{page}=2 
  \pagenumbering{arabic} 
  \setcounter{page}{2} 
  \fi
}

\title{One in Eight OpenAlex Abstracts Has Integrity Issues}
\runningtitle{One in Eight OpenAlex Abstracts Has Integrity Issues}

\author{
  Seorin Kim\textsuperscript{1,*} \\ 
  \orcidlinkc{0009-0001-5086-2704} \\
  \And
  Vincent Holst \textsuperscript{1} \\
  \orcidlinkc{0009-0002-4117-4966} \\
  \And
  Vincent Ginis \textsuperscript{1,2} \\ 
  \orcidlinkc{0000-0003-0063-9608} \\
  \and
  \textsuperscript{1}Data Analytics Lab, Vrije Universiteit Brussel, 1050, Brussel, Belgium \\ 
  \textsuperscript{2}School of Engineering and Applied Sciences, Harvard University, Cambridge, Massachusetts, 02138, USA
}
    
\begin{document}
\maketitle
\renewcommand{\thefootnote}{} 
\footnotetext[1]{*\,
Corresponding author: \href{mailto:seorin.kim@vub.be}{seorin.kim@vub.be}}
\renewcommand{\thefootnote}{\arabic{footnote}} 
\thispagestyle{plain} 

\begin{abstract}
Scientific abstracts are increasingly used as primary data in computational metascience research, yet the quality of these abstracts in widely used bibliographic databases has not been systematically examined. We assess the integrity of 10,000 randomly sampled English-language journal abstracts from OpenAlex using a two-stage annotation protocol combining human expert review and large language model classification. We identify seven distinct failure modes and find that 12\% of abstracts have integrity issues, with insufficient content and misplaced metadata being the most prevalent. We discuss implications for downstream research and describe a forthcoming community portal to support collective annotation efforts.
\end{abstract}


\section{Introduction}
The past decade has seen a rapid expansion in computational approaches to the study of science. Scientific abstracts have emerged as a key input for a growing range of analytical pipelines. Applications span from embedding-based novelty measurement \parencite{Arts2025BeyondCitations} to semantic knowledge graph construction \parencite{Tosi2021SciKGraph} to LLM-based pipelines that extract structured attributes such as research questions, dataset types, and methodological choices from publication text \parencite{kim2026turningcitationnetworksinside}. Common to all these approaches is the assumption that the text stored in the abstract field of bibliographic records is a valid abstract of the corresponding paper.

OpenAlex \parencite{priem2022openalexfullyopenindexscholarly}, the most widely used open bibliographic database, is the predominant data source for such analyses. Launched in 2022 as a fully open successor to Microsoft Academic Graph, OpenAlex now indexes over 250 million scholarly works and exposes a freely accessible API. 

Despite its large availability, \textcite{Culbert2025Reference} have shown that OpenAlex had lower abstract coverage (87\%) compared to Web of Science and Scopus (over 92\%), through a systematic comparison of OpenAlex coverage against Web of Science and Scopus with 16.8 million papers. \textcite{Alonso2025Coverage} further shed light on the completeness and accuracy of metadata for African-journal publications and identified issues with incomplete abstracts and non-abstract content in the abstract field in a manual validation of 60 entries. Despite prior work, no study has examined, at scale, the prevalence of abstract integrity failures in which OpenAlex indicates that an abstract exists (\texttt{has\_abstract=True}), but the stored text is not a valid scientific abstract. 

This gap is important because the integrity of the abstract text directly determines the validity of downstream analyses. When a novelty score is computed from the embedding of what is in fact a bibliographic citation string, or when a topic model ingests a cookie banner as scientific content, the failure propagates silently into substantive conclusions about how science works. As abstract-based methods grow in sophistication and influence, the assumption of data integrity becomes an increasingly load-bearing one.

We address this gap through a systematic, two-stage assessment of abstract integrity across 10,000 randomly sampled OpenAlex records. Our contributions are: (1) a taxonomy of seven abstract integrity failure modes grounded in multi-annotator deliberation; (2) an empirical estimate of the prevalence of each failure mode in a representative sample; (3) a validated LLM-based classification prompt that achieves 96\% agreement with human ground truth; and (4) a publicly released annotated dataset and an ongoing community initiative to extend coverage across the full OpenAlex corpus.

\section{Data and Methods} \label{sec:data_method}

\subsection{Sampling}
We drew a random sample of 10,000 English-language journal articles from the OpenAlex snapshot (11-11-2025). Eligible records met the following criteria: publication type is journal article; language is English; \texttt{has\_abstract} is True; the record is not marked as retracted; \texttt{cited\_by\_count} is at least 1; and publication year falls between 1900 and 2025. Here, restricting to cited records ensures practical relevance as uncited, recent, or retracted records are less likely to feed into active research pipelines. This restriction also means that our estimated failure rate is a conservative lower bound, and records outside these filters may exhibit higher failure rates.

\subsection{Stage 1: Human and LLM Annotation of 1,000 Abstracts}
The first 1,000 abstracts were independently rated by four annotators: two human experts and two LLM-based annotators, implemented through Claude Code using Claude Opus 4.6 subagents and OpenAI Codex using GPT-5.4 Codex subagents. Each annotator rated each abstract as valid or invalid according to the question, ``\textit{Is this a complete and meaningful scientific abstract?}''

\begin{figure}[!tbh]
    \centering
    \includegraphics[width=\linewidth]{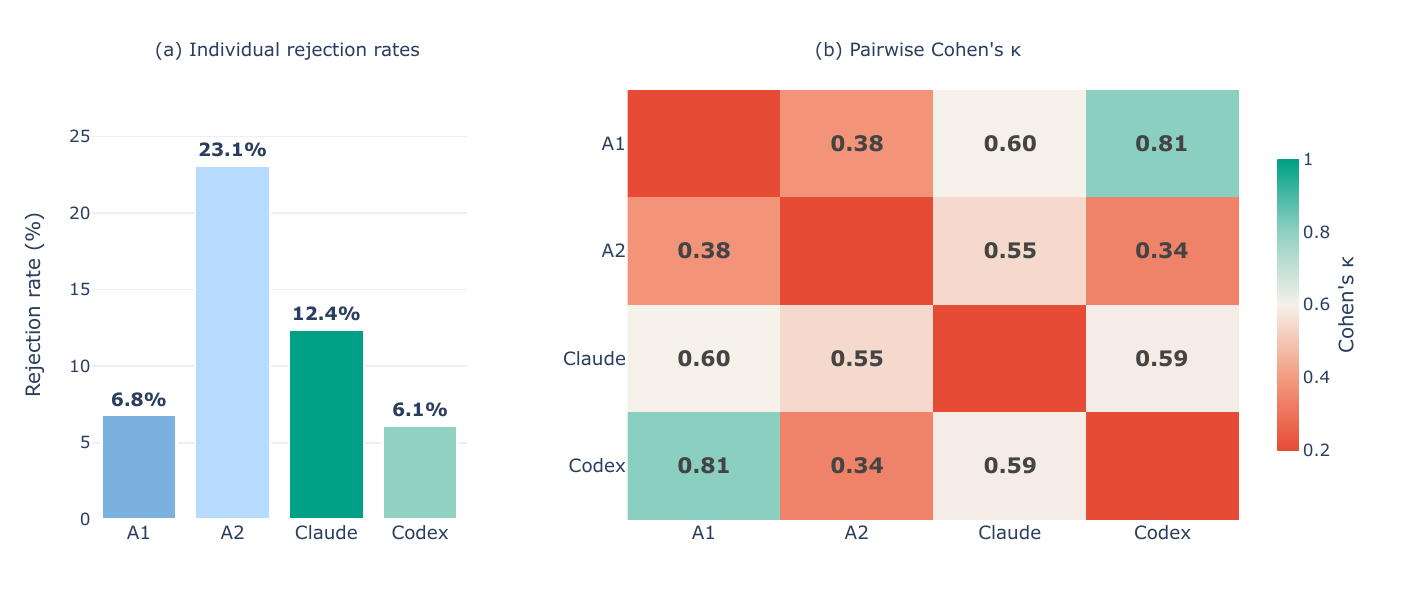}
    \caption{Inter-annotator agreement across four annotators (Human: A1 and A2; LLM: Claude and Codex). (a) Individual rejection rates. (b) Pairwise Cohen's $\kappa$. Agreement does not cluster by annotator type.}
    \label{fig:agreement}
\end{figure}

Rejection rates varied considerably across annotators. Annotator 2 was the strictest at 23.1\%, followed by Claude (12.4\%), Annotator 1 (human, 6.8\%), and Codex (6.1\%). Overall inter-annotator agreement measured by Fleiss' $\kappa$ was 0.50 (moderate). Pairwise Cohen's $\kappa$ ranged from 0.34 (Annotator 2 vs. Codex) to 0.81 (Annotator 1 vs. Codex), with no consistent pattern of higher agreement within human pairs or within LLM pairs (Figure~\ref{fig:agreement}). Notably, the two human annotators exhibited the second-lowest pairwise agreement ($\kappa = 0.38$), indicating that the disagreement structure did not align with annotator type.

Table~\ref{tab:results} shows the distribution of four-way voting patterns. Of the 1,000 abstracts, 804 received unanimous ratings (YYYY: 752; NNNN: 52). Among the 196 entries that disagreed among the four annotators (hereafter, disagreed entries), the dominant pattern was YNYY (n = 115), in which Annotator 2 alone rejected the abstract, consistent with their overall strictness. The pattern most suggestive of a human-LLM divide (NNYY: both humans reject, both LLMs accept) occurred only 5 times, providing little evidence for a systematic human-machine split.

The two human annotators then engaged in structured deliberation over all disagreed entries to resolve labels and establish boundary rules. Key decisions included that (1) short abstracts are valid if they convey both methods and results; (2) case reports are valid regardless of structure; (3) HTML markup around otherwise valid content is not a failure. The deliberation also surfaced systematic patterns, including PubMed entries returning only the conclusion of structured abstracts, and truncations at exactly 200 characters, suggesting an ingestion-level character limit.

\begin{table}[!thb]
\centering
\label{tab:results}
\caption{Four-way voting patterns across 1,000 abstracts ( A1 = Annotator 1, A2 = Annotator 2, C = Claude, X = Codex). 75\% of the assessed abstracts are valid by the four judges. A2 appeared to be the most strict, as 11.5\% are only rejected by it.}
\begin{tabular}{|c|c|c|c|c|c|c|}
\hline
\textbf{A1}               & \textbf{A2}               & \textbf{C}                & \textbf{X}                & \textbf{Count} & \textbf{\% of total} & \textbf{Note}                          \\ \hline
Y                         & Y                         & Y                         & Y                         & 752            & 75.2\%               & Unanimous valid                        \\ \hline
Y                         & \cellcolor[HTML]{EFEFEF}N & Y                         & Y                         & 115            & 11.5\%               & A2 only rejects                        \\ \hline
\cellcolor[HTML]{EFEFEF}N & \cellcolor[HTML]{EFEFEF}N & \cellcolor[HTML]{EFEFEF}N & \cellcolor[HTML]{EFEFEF}N & 52             & 5.2\%                & Unanimous invalid                      \\ \hline
Y                         & \cellcolor[HTML]{EFEFEF}N & \cellcolor[HTML]{EFEFEF}N & Y                         & 43             & 4.3\%                & A2 \& Claude reject                    \\ \hline
Y                         & Y                         & \cellcolor[HTML]{EFEFEF}N & Y                         & 14             & 1.4\%                & Claude only rejects                    \\ \hline
\cellcolor[HTML]{EFEFEF}N & \cellcolor[HTML]{EFEFEF}N & \cellcolor[HTML]{EFEFEF}N & Y                         & 9              & 0.9\%                & Only Codex accepts                     \\ \hline
Y                         & \cellcolor[HTML]{EFEFEF}N & \cellcolor[HTML]{EFEFEF}N & \cellcolor[HTML]{EFEFEF}N & 6              & 0.6\%                & Only A1 accepts                        \\ \hline
\cellcolor[HTML]{EFEFEF}N & \cellcolor[HTML]{EFEFEF}N & Y                         & Y                         & 5              & 0.5\%                & Both humans reject, both LLMs   accept \\ \hline
Y                         & Y                         & Y                         & \cellcolor[HTML]{EFEFEF}N & 2              & 0.2\%                & Only Codex rejects                     \\ \hline
\cellcolor[HTML]{EFEFEF}N & Y                         & Y                         & Y                         & 1              & 0.1\%                & Only A1 rejects                        \\ \hline
\cellcolor[HTML]{EFEFEF}N & \cellcolor[HTML]{EFEFEF}N & Y                         & \cellcolor[HTML]{EFEFEF}N & 1              & 0.1\%                & Only Claude accepts                    \\ \hline
\end{tabular}
\end{table}

\subsection{Failure Mode Taxonomy}
From the rejected entries in the 1,000-abstract sample, we derived a taxonomy of seven non-overlapping failure modes.

\paragraph{Insufficient abstract content.} Abstracts that contain only a conclusion snippet, a bare research question, or a repetition of the paper title without substantive content. 

\begin{quote}
\textbf{Paper ID:} https://openalex.org/W2467617132 \\
\textbf{Abstract:} In this series 17.34 \% patients developed locoregional recurrence for mean follow-up duration of 3.5 years. Mean disease-free interval was 20.52 months. Lymph node involvement had significant correlation with LRR. \\
\textbf{Why rejected:} Statistical findings are reported, but there is no indication of what patient population, disease, or treatment was studied. Only the conclusion was scraped from PubMed.
\end{quote}

\paragraph{Bibliographic / repository metadata.} Citation strings, DOI-only entries, or tables of contents instead of prose. 
\begin{quote}
\textbf{Paper ID:} https://openalex.org/W2897795388 \\
\textbf{Abstract:} Submitted by Natalia Cristina Aragao Gomes (ntgomes@stj.jus.br) on 2015-03-25T19:39:20Z No. of bitstreams: 1 provas\_processo\_ambiental\_greco.pdf: 558295 bytes, checksum: 18c54ebbbe5a3db5d320fc8a02f16c9a (MD5) \\
\textbf{Why rejected:} DSpace repository submission metadata (timestamp, filename, checksum). Not an abstract.
\end{quote}

\paragraph{Wrong document section.} An introduction, body paragraph, or other non-abstract section appears in place of the abstract. 
\begin{quote}
\textbf{Paper ID:} https://openalex.org/W1509045375 Annotator comment: "Story telling" \\
\textbf{Abstract:} THE FOREIGN LANGUAGE TEACHING PROFESSION, more than any other it seems to me, is in perpetual search for new slogans.' I recall the comment of a colleague back in 1971, when I was a newcomer to the profession and interested in exploring the concept of communicative competence.2 Communicative competence, he said to me, that'll be a good topic for a year or two. Then what are you going to do? My colleague did not foresee the exploration and discussion that would ensue. Today, more than a decade later, interest in the concept of communicative competence has not only not waned, it continues to grow and has led to the elaboration of descriptive models that have in turn provided frameworks for further research into the nature and acquisition of second-language proficiency.3 The pre-eminence of communicative competence as a focus of discussions of second-language teaching and evaluation was nowhere more apparent than at the October 1984 TOEFL Invitational Conference at the Educational Testing Service in Princeton. The conference was held to consider revision \\
\textbf{Why rejected}: First-person narrative prose from the introduction of an academic paper, with footnotes and personal anecdotes. Not an abstract.

\end{quote}

\paragraph{Web-scrape artefacts.} Cookie banners, navigation menus, error messages, or other web interface content captured during HTML-based ingestion.
\begin{quote}
\textbf{Paper ID:} https://openalex.org/W1975905332 \\
\textbf{Abstract:} ADVERTISEMENT RETURN TO ISSUEPREVArticleNEXTC-H and C-D Bonds: An Experimental Approach to the Identity of C-H Bonds by Their Conversion to C-D BondsAlex T. Rowland View Author Information Department of Chemistry, Gettysburg College, Gettysburg, PA 17325Cite this: J. Chem. Educ. 2003, 80, 3, 311 [...] SUBJECTS:Acid and base chemistry,Hydrogen,Hydrogen isotopes,Nuclear magnetic resonance spectroscopy,Reactivity Get e-Alerts \\
\textbf{Why rejected:} Another ACS Publications scrape: article navigation, author info, citation metadata, altmetrics, and subject keywords. Not an abstract.

\end{quote}

\paragraph{Truncated abstracts.} Text cut off mid-sentence, suggesting an ingestion-level character limit or extraction failure.
\begin{quote}
\textbf{Paper ID:} https://openalex.org/W2791313856 \\
\textbf{Abstract:} Informal credit was an important sector of financial activity since middle age. The work analyses a specific credti activity developed within Mediterranean by the Redenzione de Captivi \\
\textbf{Why rejected:} Ends mid-phrase ("the Redenzione de Captivi"); the abstract is cut off. At 184 characters, this is likely a character-limit truncation.
\end{quote}

\paragraph{No abstract / placeholders.} Strings such as "Abstract not available," "N/A," or blank fields despite has\_abstract=True.
\begin{quote}
\textbf{Paper ID:} https://openalex.org/W2171347833 \\
\textbf{Abstract:} Abstract ChemInform is a weekly Abstracting Service, delivering concise information at a glance that was extracted from about 100 leading journals. To access a ChemInform Abstract of an article which was published elsewhere, please select a "Full Text" option. The original article is trackable via the "References" option. \\
\textbf{Why rejected:} A boilerplate ChemInform service description, identical across thousands of entries. No paper-specific content.

\end{quote}

\paragraph{Wrong scholarly genre.} Content from a non-journal-article document type (e.g., a book chapter summary or conference program entry) that has been misclassified.
\begin{quote}
\textbf{Paper ID:} https://openalex.org/W2594187833 \\
\textbf{Abstract:} An article that appeared in issue 62(4) [1] had the first and second authors' names incorrectly displayed in the HTML.The first and last names of Kathy Ruble and Anna George were reversed.We apologize for the error. \\
\textbf{Why rejected:} An erratum notice correcting an author-name display error. No research content.

\end{quote}

\begin{figure}[!tbh]
    \centering
    \includegraphics[width=0.5\linewidth]{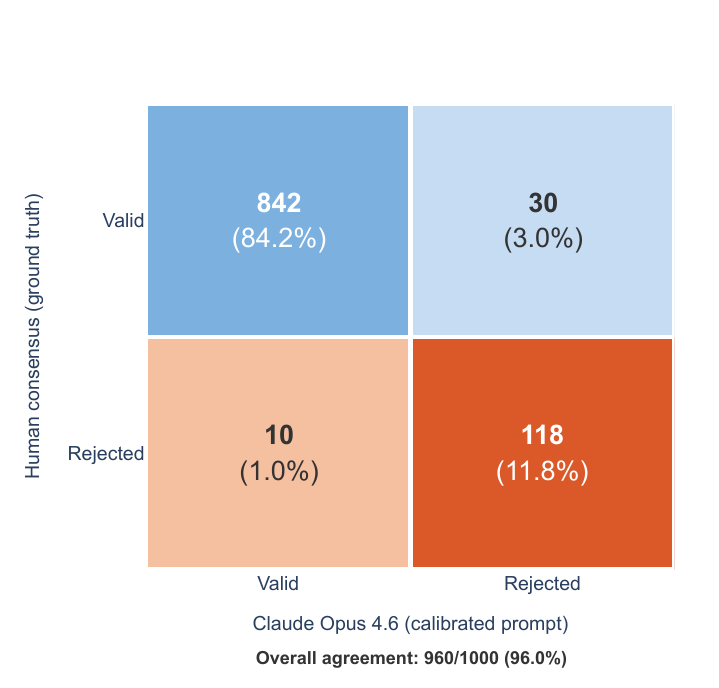}
    \caption{Binary confusion matrix comparing Claude Opus 4.6 (calibrated prompt) against the human consensus ground truth on the first 1,000 abstracts (96.0\% agreement) after the annotators' deliberation.}
    \label{fig:confusion}
\end{figure}
\subsection{Stage 2: LLM Classification of 10,000 Abstracts}

Using insights from the annotators' deliberation, we calibrated a classification prompt for inference with a large language model. The prompt encodes the failure-mode taxonomy and the key boundary decisions made during deliberation. We applied this prompt to all 10,000 abstracts using Claude Opus 4.6 via batch inference. The complete prompt is provided in the Appendix \ref{sec:app_prompt}.

To validate the prompt, we evaluated it on 1,000 abstracts using the previously discussed human-labeled ground truth. Overall binary agreement was 96.0\%. As shown in Figure~\ref{fig:confusion}, of the 40 disagreements, 30 were false positives, where Claude rejected an abstract that the human consensus accepted, and 10 were false negatives, where Claude accepted an abstract that the human consensus rejected. The higher false-positive rate suggests that the calibrated prompt was slightly more conservative than the human consensus in boundary cases. Note that the quality of the prompt was tested on the same set of abstracts used in the initial annotation for calibration. However, the risk of overfitting is minimal, as the calibration primarily involved deriving seven failure modes and qualitative boundary rules rather than supervised label fitting, and the model was not fine-tuned on any labeled data.

\begin{figure}[!tbh]
    \centering
    \includegraphics[width=0.7\linewidth]{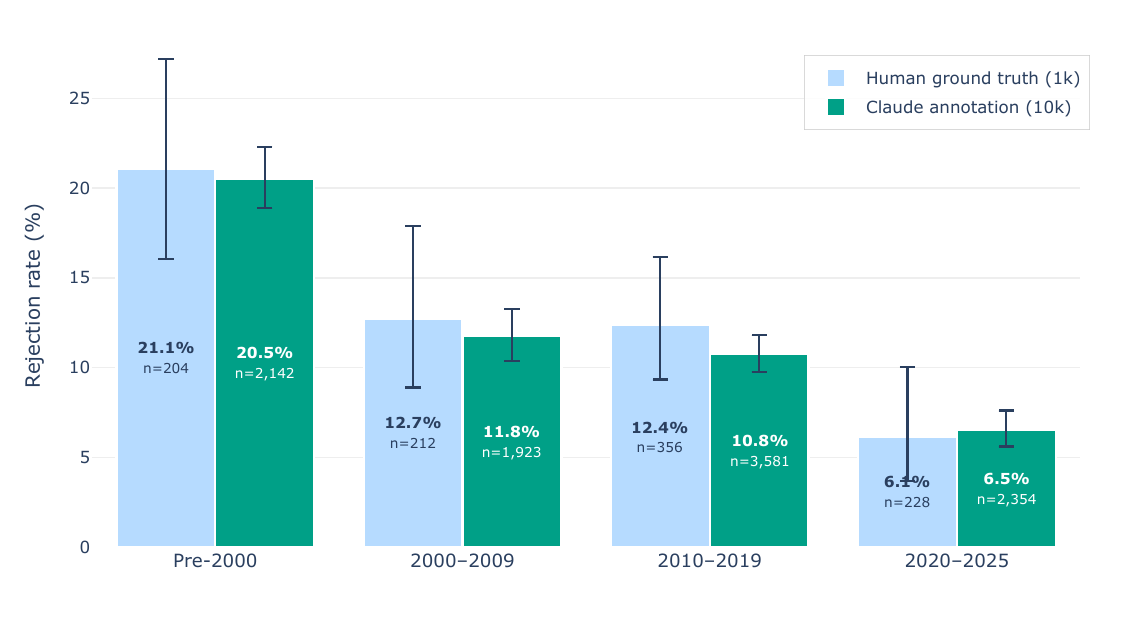}
    \caption{Rejection rate by publication period. Rejection rates declined consistently across publication periods for both the human ground truth (1k, after annotator deliberation) and the Claude annotation (10k), and the two sources agreed closely in every period. }
    \label{fig:rejection}
\end{figure}

Figure~\ref{fig:rejection} demonstrates the declining rejection rates across publication periods for both the human ground truth (1,000 papers) and the Claude annotation (10,000 papers). The two sources agreed closely across all periods. Abstracts published before 2000 had the highest rejection rate, 21.1\% by human annotators (n=204) and 20.5\% by Claude (n=2,142), with wide confidence intervals reflecting smaller and noisier pre-digital samples. Rates dropped to approximately 12\% for the 2000–2009 and 2010–2019 periods, then fell sharply to around 6\% for 2020–2025, suggesting that more recent deposits on OpenAlex are substantially cleaner.

\section{Results}

\begin{figure}[!tbh]
    \centering
    \includegraphics[width=0.8\linewidth]{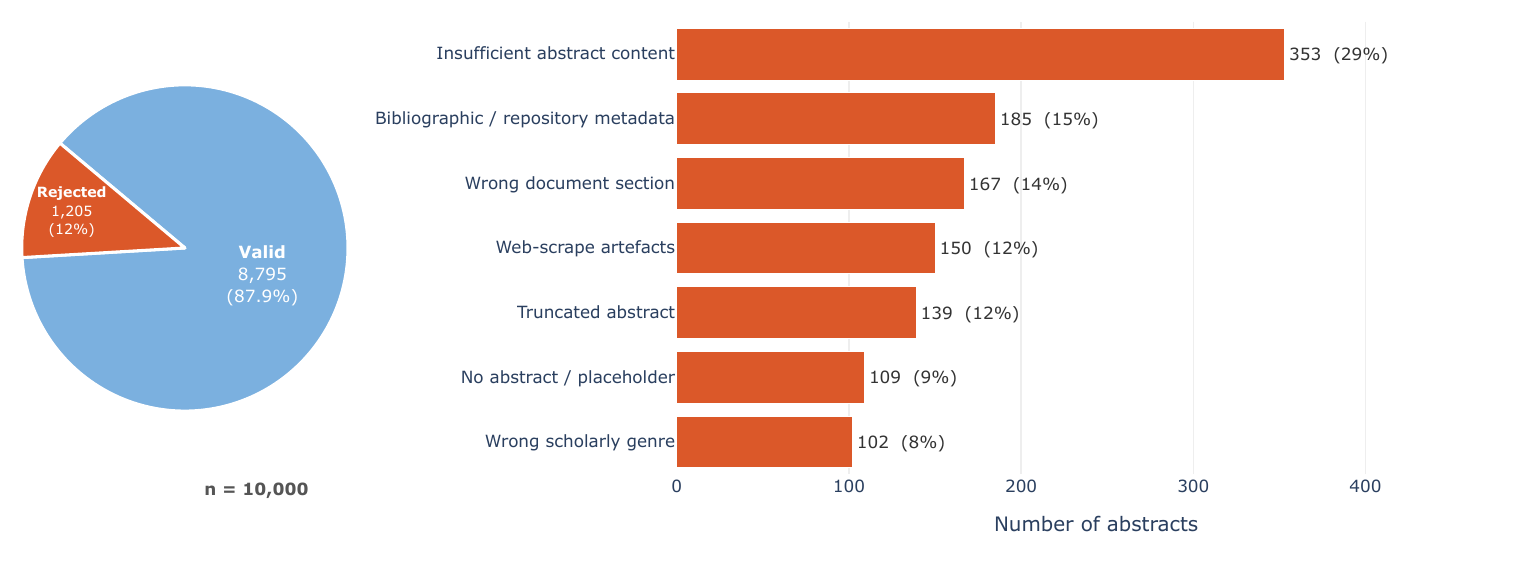}
    \caption{Distribution of abstract integrity failures across 10,000 OpenAlex abstracts. The inset shows the overall valid/rejected split. Seven failure modes are shown, along with their absolute counts and percentages of all rejected entries.}
    \label{fig:failure_modes}
\end{figure}

Across all 10,000 abstracts, 12\% were flagged as having integrity issues. Figure~\ref{fig:failure_modes} shows the distribution of failure modes among flagged entries. The most prevalent failure mode is insufficient abstract content (29.3\% of flagged entries). This category includes conclusion-only snippets characteristic of PubMed structured abstracts returned in incomplete form, bare research questions without accompanying findings, and entries that merely restate the paper title. The second most common mode is bibliographic or repository metadata (15.4\%), in which the abstract field contains citation strings, DOI-only entries, or tables of contents. The wrong document section accounts for 13.9\% of failures.
The remaining categories are less prevalent but diagnostically important. Web-scrape artefacts constitute 12.4\% of failures. Truncated abstracts account for 11.5\%; many are cut off at exactly 200 characters, consistent with a character-limit cap at the ingestion stage. Missing abstract placeholders account for 9\% of failures, and wrong scholarly genre accounts for the remaining 8.5\%.

\begin{figure}[!tbh]
    \centering
    \includegraphics[width=0.7\linewidth]{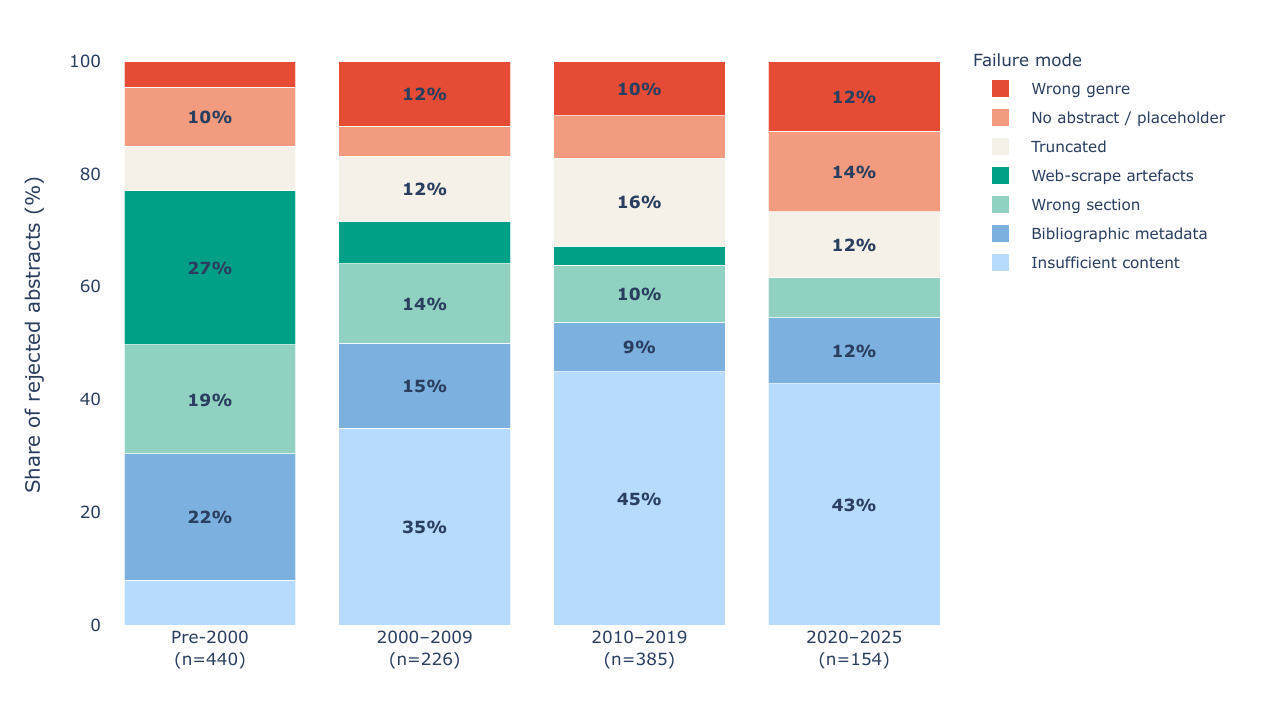}
    \caption{Failure mode composition per period across 10,000 OpenAlex abstracts.}
    \label{fig:failure_by_period}
\end{figure}

Investigating patterns across publication periods (Figure~\ref{fig:failure_by_period}), the mix of failure modes shifted markedly, revealing different underlying causes of abstract quality issues at different points in time. Web-scrape artefacts were the dominant failure mode in the pre-2000 corpus (27\% of rejections), likely reflecting early digitization practices and inconsistent OCR or HTML extraction. This aligns with the fact that bibliographic metadata contamination is the highest in the pre-2000 corpus (22\% of rejections). 
From 2000 onward, insufficient abstract content emerged as the leading category, growing from 35\% of rejections in 2000–2009 to 43–45\% in the two most recent periods. This may reflect the aforementioned errors in PubMed-derived records, in which only the conclusion section of the abstract is entered as the abstract. Lastly, the rate of rejections for truncated text increased from 8\% in pre-2000 to 16\% in 2010-2019, indicating that the 200-character truncation pattern is a systematic error rather than an incidental ingestion issue. 

Considering that our sample is restricted to English-language, cited journal articles published before 2025, these results represent a lower bound on the overall failure rate in OpenAlex. Non-English records, uncited papers, preprints, and very recent publications likely exhibit higher failure rates.

\section{Discussion and Conclusions}
This study provides the first large-scale, systematic assessment of abstract integrity in OpenAlex. The finding that approximately 12\% of abstracts exhibit integrity issues, even in a conservative, filtered sample, has direct implications for computational metascience research.

The consequences of abstract integrity failures depend on the downstream application. Embedding-based methods are particularly sensitive. For instance, a novelty score computed from the embedding of a metadata string rather than a scientific abstract introduces noise that is invisible to the researcher and difficult to detect post hoc. Topic models and LLM-based extraction pipelines face analogous risks when web-scrape artefacts or wrong-section content are included as inference data. Because failures are not uniformly distributed across domains, time periods, or publication sources, they may introduce systematic biases rather than random noise. In particular, those biases are difficult to diagnose without direct inspection of the abstract text.

Our taxonomy of failure modes offers a practical starting point for researchers who wish to filter their own datasets. The calibrated classification prompt we release can be applied directly to any OpenAlex-derived abstract corpus. Researchers working at scale may also use the annotated 10,000-entry dataset to train or evaluate purpose-built classifiers.

Beyond this initial release, we would like to initiate a discussion with the science-of-science community on how to collectively address the issue of abstract integrity at scale. At this stage, two broad responses seem possible to us. First, systematic cleaning of compromised records at the database level, and second, principled awareness, in which researchers filter their own pipelines and report their cleaning decisions as part of their methodology. We believe that building consensus around either path is a collective infrastructure challenge that no single research group should face alone. As a concrete first step, we release our annotated dataset, failure-mode taxonomy, and calibrated classification prompt as open resources, and invite the community to extend, correct, and challenge them. Finally, we urge the community to move from the invisible assumption of abstract integrity to a documented property of the data we work with. As our experiment has shown, recent advances in LLMs and agentic AI make a qualitative review of inherently noisy bibliometric data tractable at scale for the first time, and the standards we hold this data to should rise accordingly.

\newpage

\section{Science Practices}
The annotated dataset (1,000 human-curated entries with four-annotator votes and deliberation records, plus 10,000 LLM-annotated entries with failure-mode labels) and the calibrated classification prompt will be publicly released. The sampling procedure and annotation guidelines will be documented in the repository (\href{https://github.com/SereneKim/abstract_integrity}{https://github.com/SereneKim/abstract\_integrity}). The LLM inference was conducted using a commercial API; the exact model versions are reported in the methods section to support reproducibility. Raw annotation outputs from all four annotators will be included to allow independent reanalysis of inter-annotator agreement.

\section{Acknowledgments}
We would like to thank the OpenAlex team for maintaining a large-scale bibliometric database that is freely accessible to all researchers. Open bibliometric databases are essential for ensuring reproducibility in science, and we hope that our investigation of abstract quality will help improve the database.

\section{Author Contributions}
SK and VH contributed equally to this work. SK: Conceptualization, methodology, writing; VH: Analysis, methodology, writing; VG: Supervision, review.

\section{Competing Interests}
The authors declare no competing interests.

\section{Funding Information}
VG acknowledges funding (IBOF-23078) from Vrije Universiteit Brussel (vub.be). VG also acknowledges support from Research Foundation Flanders under Grant No.G032822N and G0K9322N. No funder played any role in the study design, data collection and analysis, decision to publish, or preparation of the manuscript.

\printbibliography

\appendix
\section*{Appendix}
\section{LLM Prompt for Classification by Failure Modes} \label{sec:app_prompt}
The classification prompt was derived from the structured resolution of 196 disagreements among four independent annotators (two human, two LLM) on a calibration set of 1,000 abstracts. During this discussion phase, recurring boundary cases were identified and codified as explicit decision rules; for instance, that short abstracts are valid provided they convey both methods and results, that case reports should not be penalised for non-standard structure, and that HTML markup around otherwise valid content does not constitute a failure. The resulting prompt achieved 96.0\% agreement with the human-consensus ground truth on the same 1,000 entries, and was then applied to annotate the full 10,000-abstract dataset.

\begin{tcolorbox}[
    colback=gray!10,
    colframe=black,
    boxrule=0.5pt,
    sharp corners,
    width=\textwidth,
    left=6pt,
    right=6pt,
    top=6pt,
    bottom=6pt,
    breakable
]

\textbf{Task}

You are given a scientific paper’s title and abstract as extracted from OpenAlex. Decide whether the abstract is a valid, usable scientific abstract or whether it fails in one of the ways listed below.

\vspace{0.6em}

\textbf{Instructions}

Respond with one of the following labels:
\vspace{0.6em}

\begin{description}[
    leftmargin=4em,
    labelindent=2em,
    labelsep=0.5em,
    itemsep=0.5em,
    font=\normalfont
]

\item[Valid] - the text is a usable scientific abstract.

\item[Web-scrape artefacts] - the text is publisher or platform webpage content (navigation, paywall, article-listing boilerplate, leaked HTML/XML tags, or encoding residue) rather than the abstract itself.

\item[No abstract / placeholder] - the record contains no substantive abstract text: a stub, an explicit “no abstract available” statement, a language placeholder like “[in Japanese]”, a registration identifier, or another minimal stand-in.

\item[Bibliographic / repository metadata] - the text is structured record information rather than prose: repository metadata, citation strings, DOI-only entries, author and affiliation lists, conference or book headers, ISBN-bearing references, or tables of contents.

\item[Wrong document section] - the text is coherent scholarly prose but from the wrong part of the document (an introduction, discussion, body passage, essay section, or foreword) rather than the abstract.

\item[Truncated abstract] - the text is recognizably a genuine abstract but has been cut off at the beginning or end, often mid-sentence or mid-word, so it is incomplete as a standalone summary.

\item[Insufficient abstract content] - the text is too brief, narrow, or uninformative to function as a meaningful standalone abstract: a conclusion-only snippet that provides no indication of methods or materials used; a bare question; a title repetition; a teaser sentence; or a minimal topic announcement.

\item[Wrong scholarly genre] - the text is coherent writing from a non-abstract scholarly genre: an editorial, erratum, letter to the editor, conference report, news piece, or similar.
\end{description}

\textbf{Guidance on borderline cases}
These rules reflect how human annotators resolved disagreements on 196 borderline abstracts.

\begin{description}[
    leftmargin=4em,
    labelindent=2em,
    labelsep=0.5em,
    itemsep=0.5em,
    font=\normalfont
]
\item[\textbf{\textit{Short abstracts}}] \hfill\\[0.1em]
A short abstract (even 1–3 sentences) is \textbf{Valid} as long as it conveys both what was studied or done (method, material, system, technique) and what was found or concluded. Brevity alone is not a reason to reject. Examples of valid short abstracts:
\begin{quote}
    \textit{“Samarium trichloride catalyzed one-pot synthesis of dihydropyrimidines has been developed. The methodology was successfully applied to various aldehydes.” (method + result)}  \vspace{0.6em} \\
    \textit{“We propose a MIMO ultraviolet communication system which can increase the data rate effectively. The BER performance of a 2×2 MIMO scheme is measured and analyzed.” (system + measurement)}
\end{quote}

\vspace{0.6em}
An abstract is \textbf{Insufficient} abstract content only when it provides a finding or conclusion with no indication of methods or materials, or when it is a bare question, title repetition, or teaser with no informational content. Examples:
\begin{quote}
    \textit{“Patients with COPD hospitalized with COVID-19 experienced increased HCU post-discharge compared to patients without COPD.” (finding only, no method/material)}  \vspace{0.6em} \\
    \textit{“How can adapting to different learning styles increase confidence in trainees?” (bare question)} \\
“A novel technology for quantifying hormone secretion from tissues, with a single-cell resolution.” (teaser, no substance)

\end{quote}

\item[\textbf{\textit{Case reports and non-standard formats}}]\hfill\\[0.1em]
Case reports, clinical observations, and brief communications are \textbf{Valid} as long as they describe what was observed and convey scientific content. Do not penalize unconventional abstract structure.

\item[\textbf{\textit{HTML markup}}]\hfill\\[0.1em]
If the abstract contains HTML or XML wrapper tags (e.g., \texttt{<div>}, \texttt{<p>}, \texttt{<jats:p>}) but the underlying text is a valid scientific abstract, label it \textbf{Valid}. Only label it \textbf{Web-scrape artefacts }when the markup replaces or overwhelms the content (navigation chrome, paywall text, article metrics widgets).

\item[\textbf{\textit{Non-English abstracts}}] \hfill\\[0.1em]
Non-English text is not a failure mode by itself. Classify based on the underlying content defect: a Chinese abstract cut off mid-sentence is \textbf{Truncated abstract}; a Dutch table of contents is \textbf{Bibliographic / repository metadata}; “[in Japanese]” with no content is \textbf{No abstract / placeholder}.

\item[\textbf{\textit{Book and review abstracts}}] \hfill\\[0.1em]
Abstracts of books, book chapters, and review articles are \textbf{Valid} if they summarize scholarly content. A table of contents or publisher catalog entry is \textbf{Bibliographic / repository metadata}.

\item[\textbf{\textit{Output format}}]\hfill\\[0.1em]
Return a JSON object: \texttt{{"label": "<one of the 8 labels>"}}.

\end{description}
\end{tcolorbox}

\end{document}